\documentclass[sn-mathphys,Numbered]{sn-jnl}

\usepackage{graphicx}%
\usepackage{multirow}%
\usepackage{amsmath,amssymb,amsfonts}%
\usepackage{amsthm}%
\usepackage{mathrsfs}%
\usepackage[title]{appendix}%
\usepackage{xcolor}%
\usepackage{textcomp}%
\usepackage{manyfoot}%
\usepackage{booktabs}%
\usepackage{algorithm}%
\usepackage{algorithmicx}%
\usepackage{algpseudocode}%
\usepackage{listings}%
\usepackage{caption}
\usepackage{subcaption}
\usepackage{easyReview}

\begin{document}

\title[Piano with artificial partner]{If Turing played piano with an artificial partner}

\author*[1]{\fnm{Dobromir} \sur{Dotov}}\email{ddotov@unomaha.edu}
\author[2]{\fnm{Dante} \sur{Camarena}}\email{dante@transforms.ai}
\author[2]{\fnm{Zack} \sur{Harris}}\email{zack.tibia@gmail.com}
\author[3]{\fnm{Joanna} \sur{Spyra}}\email{spyraj@gmail.com}
\author[2]{\fnm{Pietro} \sur{Gagliano}}\email{pbgagliano@gmail.com}
\author[3,4]{\fnm{Laurel} \sur{Trainor}}\email{ljt@mcmaster.ca}

\affil*[1]{\orgdiv{Department of Biomechanics}, \orgname{University of Nebraska Omaha}, \orgaddress{\street{6001 Dodge Street}, \city{Omaha}}, \postcode{68182}, \state{Nebraska}, \country{USA}}

\affil[2]{\orgname{Transforms.AI}, \orgaddress{\city{Toronto}, \state{Ontario}, \country{Canada}}}

\affil[3]{\orgdiv{Department of Psychology, Neuroscience and Behaviour}, \orgname{McMaster University} \orgaddress{\street{1280 Main Street West}, \city{Hamilton}, \postcode{L8S4K1}, \state{Ontario}, \country{Canada}}}

\affil[4]{\orgdiv{LIVELab}, \orgname{McMaster University} \orgaddress{\street{1280 Main Street West}, \city{Hamilton}, \postcode{L8S4K1}, \state{Ontario}, \country{Canada}}}

\abstract{Music is an inherently social activity that allows people to share experiences and feel connected with one another. There has been little progress in designing artificial partners exhibiting a similar social experience as playing with another person. Neural network architectures that implement generative models, such as large language models, are suited for producing musical scores. Playing music socially, however, involves more than playing a score; it must complement the other musicians’ ideas and keep time correctly. We addressed the question of whether a convincing social experience is made possible by a generative model trained to produce musical scores, not necessarily optimized for synchronization and continuation. The network, a variational autoencoder trained on a large corpus of digital scores, was adapted for a timed call-and-response task with a human partner. Participants played piano with a human or artificial partner--in various configurations--and rated the performance quality and first-person experience of self-other integration. Overall, the artificial partners held promise but were rated lower than human partners. The artificial partner with simplest design and highest similarity parameter was not rated differently from the human partners on some measures, suggesting that interactive rather than generative sophistication is important in enabling social AI.}

\keywords{AI, inter-personal coordination, generative model, music, self-other integration, synchronization, togetherness}

\maketitle

\section{Introduction}\label{sec1}

The recent rise in popular interest in large language models (LLMs) has attracted considerable attention to deep neural networks as possible basis for artificial intelligence (AI). Such models are capable of taking questions about a very wide range of topics and giving reasonably informed answers in natural language. More technically, they are generative models that produce a response by estimating what would be the most probable continuation of the input prompt given the statistical regularities in a large training data set of text-based information \citep{wolfram_what_2023}.

Like language, music can be represented in a sequential format too, suggesting that generative models implemented by deep neural networks can serve as the basis for musical AI. The interest in academic research on application of AI to music has been increasing steadily for the past ten years. \footnote[1]{The citation count for peer-reviewed papers containing both "music" and "artificial intelligence" in their titles and abstracts shows a steady increase: \url{https://app.dimensions.ai/discover/publication?search_mode=content&search_text="artificial intelligence"ANDmusic&search_type=kws&search_field=text_search}.} Most of the work has been focused on generating, analyzing, and classifying music in ways that mimic human skill, and evaluating musical AI by testing whether its compositions sound like they were written by a particular composer. This is consistent with the overall modus operandi in AI research, Turing’s \emph{imitation game}, which is to design machines to mimic specific human cognitive functions \citep{turing_computing_1950}, despite historical evidence that truly groundbreaking technology often emerges accidentally and in unexpected domains \citep{ihde_technology_1999}.

This approach does not address the social and collaborative aspects of music. Throughout history and across cultures, music has most typically involved multiple people engaging in music-making together as a social behavior. Therefore, an alternative approach is to test “\emph{[i]f a musician could ‘jam’ with an unseen Jam Factory and with an unseen human musician for as long as desired and was unable to tell which was the human, then, according to the Turing test, Jam Factory would have exhibited ‘intelligence’}” \citep{belgum_turing_1988}. While generative AI models are improving immensely in their capacity to learn statistical patterns and produce visual art, speech, and music in various styles, here we focus on musical interaction between a person and a neural network, asking how the human experiences this interaction. Without diminishing the importance of deep background knowledge in enabling human social skills, we emphasize the importance of inter-personal interaction and coordination of movement.

Collaborative music-making technology has the capacity to enable a wide range of social activities because music is inherently social. The archaeological evidence dates the use of musical instruments at least 30,000 years back, while singing and drumming are probably even older \citep{conard_new_2009}. Although musical behavior can be found in individuals listening or playing alone, around the world it occurs mainly in groups ranging from duets to hundreds of participants. Coordinated action is an essential aspect of group music-making, a defining aspect of social behavior, kinship, and group survival \citep{honing_without_2015,huron_is_2001,patel_evolutionary_2014,salimpoor_anatomically_2011,savage_statistical_2015,trainor_origins_2015}, although the value of exact synchronization may vary from culture to culture and across musical contexts \citep{benadon_quantitative_2018,davies_effect_2013,lucas_inter-group_2011}. Affiliation and cooperation among people increase after they experience synchronous movement with each other \citep{hove_its_2009,mogan_be_2017,valdesolo_rhythm_2010}. Infants, without having been exposed to an exhaustive musical repertoire, show early musical preferences, social-emotional responses to music, and rate-sensitive motoric responses to musical rhythm \citep{cirelli_interpersonal_2014,cirelli_rhythm_2018,trainor_rhythm_2019,zentner_rhythmic_2010}. From an evolutionary perspective, many non-human species, engage in vocal chorusing, in which individuals may increase survival either by coordinating their chorusing (cooperating) or by trying to make their vocal signals stand out from those of others (competing) \citep{gamba_indris_2016,greenfield_evolution_2017,ravignani_chorusing_2014,ravignani_interactive_2019}. While most musical behaviors may be collective, humans also engage in individual music making, consistent with the idea that music may have evolved from the collective to the individual. This relates to the Vygotskian notion, popularized in cognitive science by Hutchins \cite{hutchins_cognition_1995}, that many cognitive skills are social processes that have become internalized. The primacy of social cognition implies that shared music performance precedes solo skills.

Musicians often report that when they are deeply engaged in group performance, they can anticipate each other's ideas and act as one. To study group music performance, aspects of inter-personal experience need to be measured. To this end, participants reported on their self-other integration \citep{aron_inclusion_1992} using an analogue visual scale rather than relying on the participants’ verbal reports of their first-person experiences and non-verbal interactions with their partners. An even stronger form of engagement with performance is described as a state of flow \citep{csikszentmihalyi_flow_1990}. Intuitively described as “losing oneself in the action”, the flow state is associated with effortless concentration, peak performance, productivity, and creativity \citep{csikszentmihalyi_flow_2014}. Importantly, this state of absorption is particularly relevant when musicians are playing together \citep{damario_judgment_2022,gaggioli_networked_2017,hart_individuality_2014}.

We designed a study to address the question whether participants enjoyed the experience of playing piano with an interactive turn-taking artificial partner. A pre-trained multi-layer neural network optimized for time sequences was selected. MusicVAE, openly accessible from Google Brain’s Magenta project, was trained on a large corpus of recorded piano performances spanning multiple eras, genres, and cultural traditions \citep{noauthor_musicvae_2018}. MusicVAE has a capacity for competent reproduction of piano melodies, achieving high musicality ratings, and has the beneficial property of smooth interpolation within its latent space \citep{roberts_hierarchical_2018}. This means that it can generate unseen sequences by sampling from a point in the latent space that has not been encountered during training but sits between two training samples. Demonstrative examples are available at \citep{noauthor_hierarchical_nodate} {and our implementation is accessible at} \citep{noauthor_aiduo}. {Two model parameters can be used to fine tune its generative capabilities. A \emph{temperature} parameter controls the strength of fluctuations within the latent space, where higher temperature allows the model to switch more freely between adjacent improvisations of the input. A \emph{similarity} parameter control how much the generative model is forced to try to imitate the input.}

We adapted MusicVAE models to play together with a human in a turn-taking paradigm with short turns. We conducted a controlled experiment where participants played piano with other participants or with the artificial partner, without blinding. The instruction was to treat the task as a practice and improvisation session and to use the timed turn-taking to exchange ideas with the partner. The imitation (similarity) and improvisation (temperature) parameters of the artificial partner were varied across trials to test participants’ responses in different conditions. Specifically, there were two levels of imitation, two levels of improvisation, and two levels of melody span. After each trial, we collected the participants’ ratings of performance quality, creativity, and whether the partner was perceived to be listening and responding in interesting ways. To be more specific about the experience of togetherness in inter-personal interaction, we collected ratings of self-other integration and flow.

\section{Methods}\label{met}

\subsection{Participants}\label{pat}

The sample of participants consisted of twenty adults ($N=20$, mean (SD) age of 33.8 (19.4) years, 17 female and 3 male). They were recruited from the department's pool of students participating for extra credit and from the laboratory's list of members of the local community who had volunteered to be experimental participants. The selection criteria included previous experience playing piano and absence of known hearing or motor disability.

\subsection{Task}\label{tas}

The score-free turn-taking task consisted of playing a short phrase on the piano, stopping to listen to the partner's response, and then resuming for another cycle without a pause in between, see Figure \ref{tas}. An individual turn lasted 8 seconds for each participant in the duo (i.e., 16 seconds for the pair). Timing was controlled for consistency by displaying a visual progress bar on the computer screen facing the participants. The full trial lasted 112 seconds, equivalent to 7 full cycles of turn-taking.

The instructions were to try to respond in a musically meaningful way to the partner, find a comfortable balance between repetition and improvisation and leading and following, and allow the musical piece to emerge on the spot from this back-and-forth performance and evolve freely. Most participants did not know each other and those who did had never player together. They did not discuss what and how they would play in advance, and avoided verbal interaction in the course of the trials. Participants were restricted to playing with one hand and without simultaneous key strokes constituting chords in order to match the monophonic output of the AI. Given the lack of score or rehearsal and the presence of additional constraints on interaction and possible musical structure, the performance consisted mostly of exchanging and developing short melodies. To prepare for playing with the AI, participants were told additionally that their partner could vary the extent to which it mimicked them or improvised, how well it could match their musical ideas, and that it was restricted to playing simple monophonic melodies.

In the human-human dyadic condition, participants sat side by side and played two separate pianos. In the human-AI dyadic condition, each participant played the same setup as in the human-human condition but the presence of the AI was indicated visually on the screen by displaying a color-coded piano roll of the most recently played keys, see Figure \ref{tas}B. {The partner was presented as a web browser application called 'AI Dynamic Duet' accessible at an address on our server} \citep{noauthor_aiduo}. The protocol was approved by the institutional review board and participants signed informed consent before performing the experiment.

\begin{figure}[t]
  \centering
  \includegraphics[width=.99\textwidth]{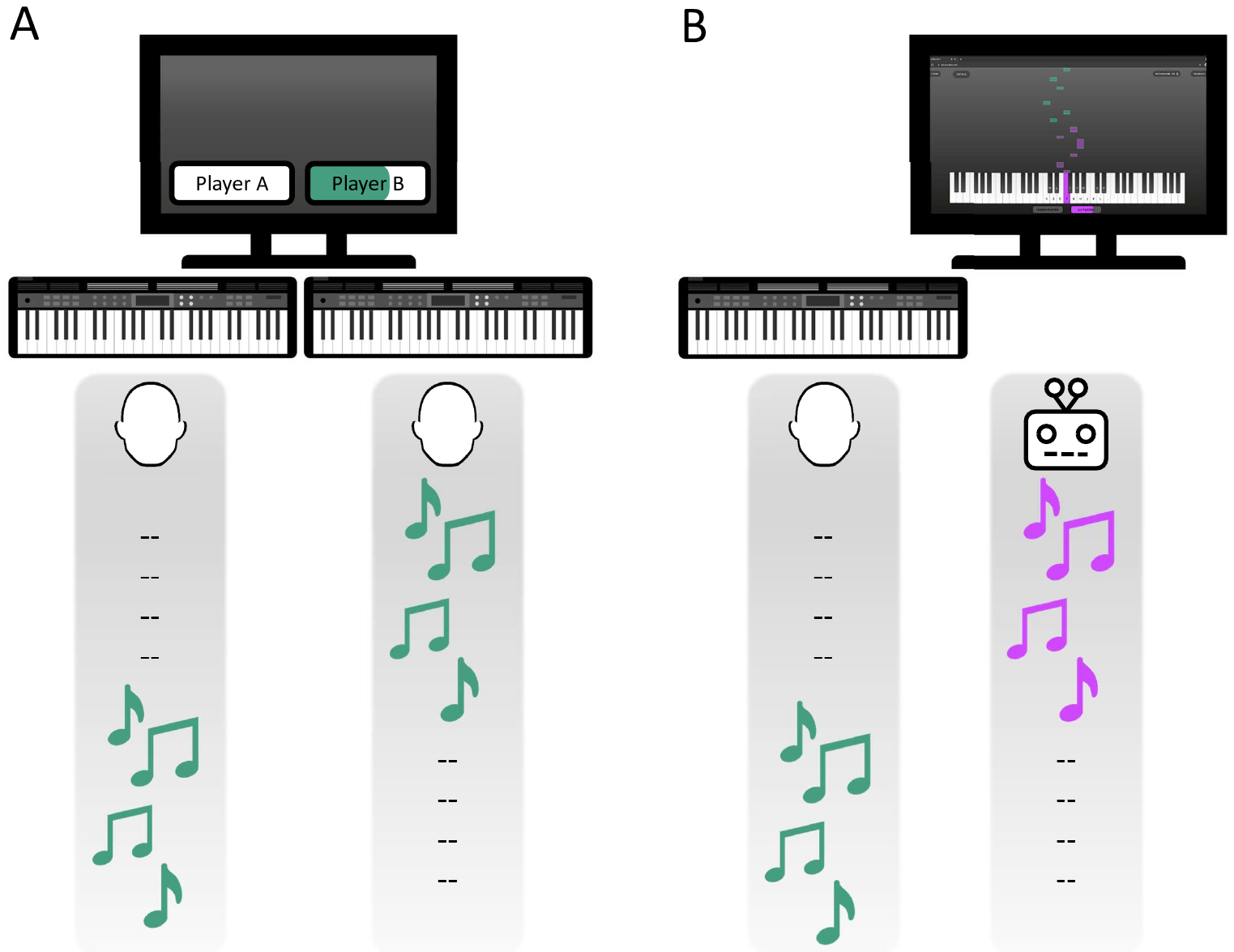}
\caption{Visual representation of the tasks where (a) the human-human duo take turns playing their individual pianos, as indicated on the screen by a diminishing coloured bar labelled Player A and Player B; and (b) human-AI turns are similarly indicated through a diminishing coloured bar, and coloured boxes representing the notes played are scrolling upwards on the screen.}
\label{tas}
\end{figure}

\subsection{Artificial Partner}\label{int}

\begin{figure}[t]
  \centering
  \includegraphics[width=.99\textwidth]{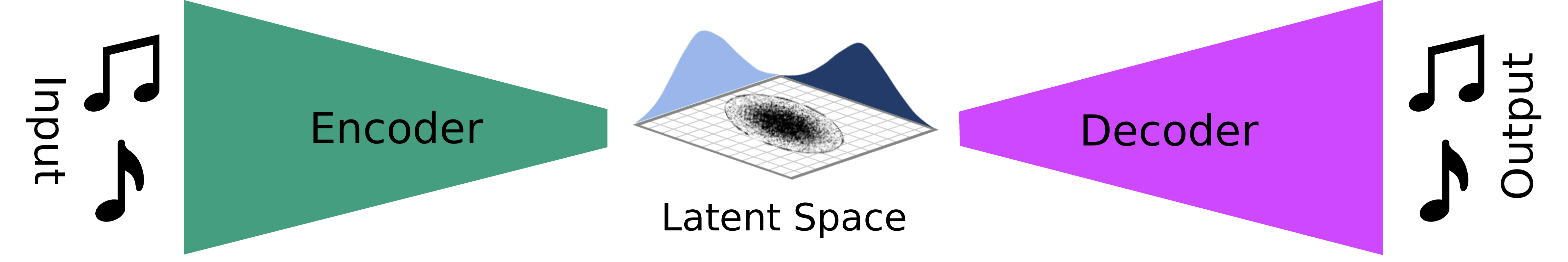}
\caption{Schematic of the architecture used to train generative models of piano note sequences. In training, the variational autoencoder is trying to reproduce the input sequences from a curated dataset by encoding them, passing them through a lower dimensional latent space with the form of a multivariate probability distribution, and then decoding them. Later, the trained generative model can be used to mimic or reconstruct input sequences as well as sample and interpolate between learned sequences.}
\label{vae}
\end{figure}

The generative modeling of time sequences is a difficult challenge that has not received as much attention as language models or fixed-size data structures such as images. The MusicVAE is a set of pre-trained deep neural networks openly available from Google Brain’s Magenta project \citep{noauthor_musicvae_2018}. They are optimized to function as generative models of time sequences, specifically of MIDI notes, not of pure sound waveforms \citep{carr_generating_2018}. While one of the main applications of machine and deep learning is to recognize and classify data structures, a generative model can be trained to reproduce time sequences. The architecture proposed by \citet{roberts_hierarchical_2018}, Figure \ref{vae}, combines RNN (LSTM networks) encoders and decoders with hierarchical properties and a low-dimensional probabilistic latent space in the middle \citep{kingma_auto-encoding_2022,rezende_stochastic_2014}. The information bottleneck in the middle of an autoencoder forces the network to extract high-level and, in theory, salient features that allow it to handle unseen corpus samples. Once trained, the network can be used to reproduce (reconstruct) or freely generate (sample) melodies in the styles of the training corpus. Furthermore, the distributional character of the latent space allows smooth interpolation. The autoencoder can interpolate between different training and input examples by activating a location in latent space that is intermediary between their corresponding locations.

While piloting the study, we found that the full version of the generative model with 16-bar input and output, albeit having the highest musical ratings \citep{roberts_hierarchical_2018}, was unsuitable for the interactive paradigm. Listening to 16-bar stretches of performance interrupted the flow of turn-taking. In the experiment, we compared two models, one trained on 2-bar and the other on 4-bar melody segments. 

The array of models in the MusicVAE set were trained on a large curated corpus of MIDI recordings \cite{roberts_hierarchical_2018}. Publicly accessible files were harvested by searching the web. Over a million unique MIDI files with 4/4 time signature were selected and were quantized to $16^{th}$ notes. These were broken into an even larger dataset of unique monophonic melodies by sectioning into 2-bars going in steps of 1 bar, excluding longer rests. This resulted in 28 million samples of 2-bar monophonic melodies. The same procedure was applied to create the 4-bar set. A smaller model with 4.4 million parameters was trained on the 2-bar set and a larger one with 11.7 million parameters on the 4-bar set.

\subsection{Apparatus}\label{app}

Participants sat at a table on performance chairs. Two professional electronic pianos with weighted keys (Roland FP80) were connected via the MIDI format to the same computer. The computer played the synthesized piano sounds through a powered studio reference monitor (Yorkville Sound YSM5) to ensure acoustic consistency between the human-human and human-AI conditions.

\subsection{Design}\label{des}

One factor of the design was the partner, consisting of a human-human condition and a human-AI condition. The second factor was the configuration of the generative model in the human-AI trials. The order of exposure to the partner condition was counter-balanced within pairs. The order of exposure to the different AI configurations within the human-AI block was randomized.

\subsection{Measures}\label{pr}

After each trial, a questionnaire collected participants' ratings of their partner and the quality of the joint performance. Each item rated a dimension of playing with the partner on a 7-point scale. The items consisted of: Musicality, Realism, Ease to interact with, Creativity and improvisation, How enjoyable was it to play together, How interesting was the last trial.

Participants also evaluated their experience after each trial. Specifically, they completed questionnaires on self-other integration (IOS) with their partner \citep{aron_inclusion_1992} and the short Flow State Scale (sFSS) \citep{jackson_long_2008}. The IOS consists of a seven-step visual scale indicating integration in terms of the amount of overlap between two circles, ranging from no integration and a score of 0 to full integration and a score of 6. The sFSS is a questionnaire consisting of nine items rating different dimensions of the state of flow. The items include questions such as "I was completely focused on the task at hand" or "I did things spontaneously and automatically without having to think". Each is rated on a five-level scale from "Strongly Disagree" to "Strongly Agree", and the nine responses are converted to an average numerical score from 1 to 5.

\subsection{Procedure}\label{pr}

Participants were assigned to pairs by order of registering and arriving to perform the experiment. Each participant performed a block of trials with a human partner and a block of trials with an artificial partner. The participant arriving earlier performed a block of human-AI trials and then a block of human-human trials with the participant arriving later. Correspondingly, the participant arriving later performed the human-human trials first and then the human-AI trials. Upon arrival and informed consent, participants were given time to familiarize themselves with the piano, with the task, and to practice playing familiar tunes from memory.

\subsection{Analysis}\label{ana}

We used linear mixed-effects modeling to compare the different configurations of the AI partner against the human-human condition. The linear-mixed effects approach benefits from the ability to account for individual variability and, importantly, can accommodate unbalanced designs \citep{singer_applied_2003}. In the present design, each participant performed one trial in each AI condition, and multiple trials in the human-human condition.

A separate model was fitted for each dependent variable, but the model specification was identical each time. It consisted of condition as a categorical predictor and random effects for the individual baseline and interaction with the predictors. The human-human condition served as baseline, technically the intercept of the regression-like model. The difference between the human-human trials and {each of the AI categories was tested in terms of the coefficient of the corresponding term in the model, see Table \ref{lmems_table}. This means that each configuration of the artificial partner was tested against the null hypothesis that its score is not different from the human-human condition.}

Two of the performance items, namely Musicality and Interesting, were excluded from analysis due to high correlation with the other items.

\section{Results}\label{rez}

\begin{figure}[hbt!]
\begin{subfigure}{.5\textwidth}
  \centering
  \includegraphics[width=.99\linewidth]{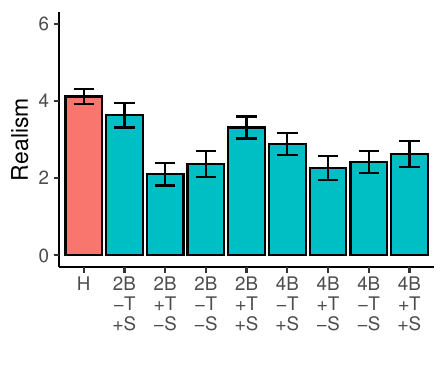}
  \caption{ }
  \label{fig3a}
\end{subfigure}
\begin{subfigure}{.5\textwidth}
  \centering
  \includegraphics[width=.99\linewidth]{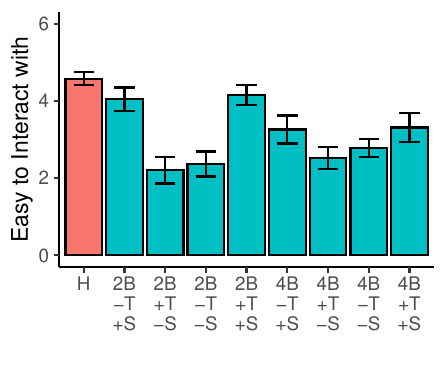}
  \caption{ }
  \label{fig3b}
\end{subfigure}

\begin{subfigure}{.5\textwidth}
  \centering
  \includegraphics[width=.99\linewidth]{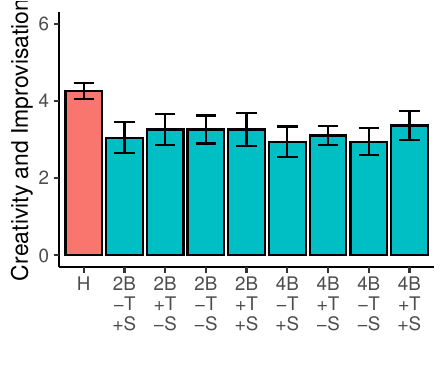}
  \caption{ }
  \label{fig3c}
\end{subfigure}
\begin{subfigure}{.5\textwidth}
  \centering
  \includegraphics[width=.99\linewidth]{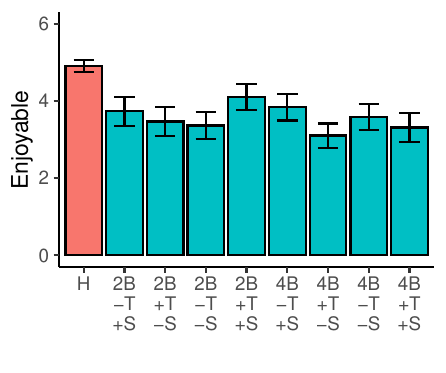}
  \caption{ }
  \label{fig3d}
\end{subfigure}
\caption{Ratings of performance quality. H: human-human performance. 2B: generative model with a two-bar time span, 4B: generative model with a four-bar time span, -T: low temperature, +T: high temperature, -S: low similarity, +S: high similarity.}
\label{perf}
\end{figure}

As Figure \ref{perf} shows, generally participants rated performance with the artificial partners lower than with a human partner. This tended to be the case for each of the eight artificial partners that was tested, but two configurations were rated similarly as humans on some measures. On \textit{Realism}, all AI partners were rated significantly lower than the baseline (i.e., human-human condition), except for the 2 bar, low temperature, high similarity (2Bar/-T/+S) condition which was not significantly different (\textit{p} = 0.202), see Table \ref{lmems_table}a and Figure \ref{fig3a}. The measure \textit{easy to interact with} was significantly lower than the baseline for six configurations, but it was not significantly different in the (2Bar/-T/+S) condition (\textit{p} = 0.154) and the 2 bar, high temperature, high similarity (2Bar/+T/+S) condition (\textit{p} = 0.288), see Table \ref{lmems_table}b and Figure \ref{fig3b}. Ratings of \textit{creativity and improvisation} were lower in five conditions and they were not significantly different from baseline in three conditions: the 2Bar/+T/-S condition (\textit{p} = 0.063), the 2Bar/+T/+S condition (\textit{p} = 0.057), and the 4 bar, high temperature, high similarity (4Bar/+T/+S) condition (\textit{p} = 0.066), see Table \ref{lmems_table}c and Figure \ref{fig3c}. 
For all artificial partners, the \textit{enjoyable} scale was lower than baseline, see Table \ref{lmems_table}d and Figure \ref{fig3d}. 

\begin{figure}[hbt!]
\begin{subfigure}{.5\textwidth}
  \centering
  \includegraphics[width=.99\linewidth]{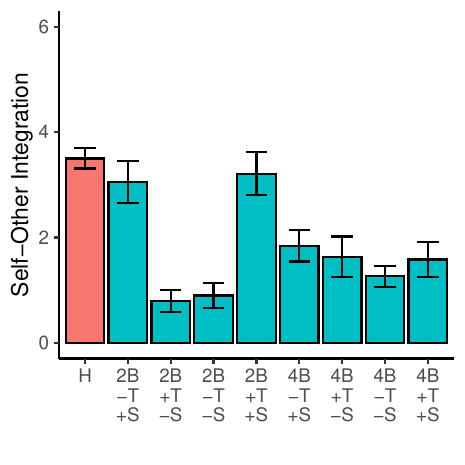}
  \caption{ }
  \label{expa}
\end{subfigure}
\begin{subfigure}{.5\textwidth}
  \centering
  \includegraphics[width=.99\linewidth]{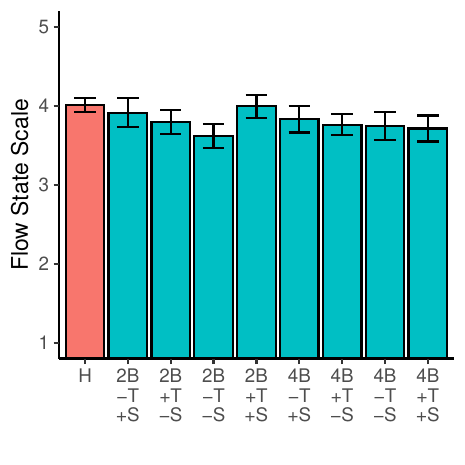}
  \caption{ }
  \label{expb}
\end{subfigure}
\caption{Ratings of experience. H: human-human performance. 2B: generative model with a two-bar time span, 4B: generative model with a four-bar time span, -T: low temperature, +T: high temperature, -S: low similarity, +S: high similarity.}
\label{exp}
\end{figure}

The measures of experience exhibited as similar pattern, with a general tendency for lower scores in artificial partner conditions, but with the same two conditions standing out, see Figure \ref{exp}. Specifically, \textit{self-other integration} was lower than baseline in six conditions, and not significantly different in two conditions: 2Bar/-T/+S (\textit{p} = 0.288) and the 2Bar/+T/+S (\textit{p} = 0.581), see Table \ref{lmems_table}e and Figure \ref{expa}. The \textit{flow state scale} tended to be lower in artificial partner than in human partner conditions, see Figure \ref{expb}, but most differences did not reach statistical significance, except for the 2 bar, low temperature, low similarity condition (\textit{p} < .05), see Table \ref{lmems_table}f.

\begin{table}[t]
\caption{Statistical models comparing configurations of the artificial partner with the performance with a human partner. The baseline in the model was the human-human condition, and the effects (standard errors in brackets) of the other conditions were evaluated in terms of how ratings changed relative to the baseline. The 8 artificial partner conditions were defined by crossing the factors of the AI, namely time span (two bars and four bars: 2B/4B), tendency to improvise (low and high temperature: -T/+T), and tendency to mimic the human input (low and high similarity: -S/+S).}
\begin{tabular*}{\linewidth}{@{\extracolsep\fill}lccc}
\toprule
  & (a) Realism & (b) Easy to interact with & (c) Creativity and Improvisation \\
\midrule

Baseline (Human)         & $4.121 \; (.347)^{***}$  & $4.169 \; (.377)^{***}$ & $4.571 \; (.288)^{***}$  \\
2Bar -Temp +Sim	         & $-.490 \; (.370)$        & $-1.116 \; (.489)^{*}$  & $-.518 \; (.349)$        \\
2Bar +Temp -Sim          & $-2.016 \; (.409)^{***}$ & $-.906 \; (.460)^{*}$   & $-2.360 \; (.487)^{***}$ \\
2Bar -Temp -Sim          & $-1.753 \; (.369)^{***}$ & $-.906 \; (.355)^{*}$   & $-2.203 \; (.440)^{***}$ \\
2Bar +Temp +Sim          & $-.805 \; (.346)^{*}$    & $-.906 \; (.448)^{*}$   & $-.413 \; (.378)$        \\
4Bar -Temp +Sim          & $-1.227 \; (.376)^{**}$  & $-1.222 \; (.486)^{*}$  & $-1.308 \; (.460)^{**}$  \\
4Bar +Temp -Sim          & $-1.858 \; (.425)^{***}$ & $-1.064 \; (.423)^{*}$  & $-2.045 \; (.437)^{***}$ \\
4Bar -Temp -Sim          & $-1.700 \; (.382)^{***}$ & $-1.222 \; (.380)^{**}$ & $-1.782 \; (.431)^{***}$ \\
4Bar +Temp +Sim          & $-1.490 \; (.423)^{***}$ & $-.801 \; (.410)$       & $-1.255 \; (.454)^{**}$  \\

\midrule
  & (d) Enjoyable & (e) Self-other Integration & (f) Flow State Scale (short) \\
\midrule

Baseline (Human)         & $5.036 \; (.296)^{***}$  & $3.472 \; (.273)^{***}$  & $3.964 \; (.159)^{***}$ \\
2Bar -Temp +Sim	         & $-1.299 \; (.397)^{**}$  & $-.419 \; (.385)$        & $-.051 \; (.146)$       \\
2Bar +Temp -Sim          & $-1.562 \; (.497)^{**}$  & $-2.683 \; (.418)^{***}$ & $-.168 \; (.131)$       \\
2Bar -Temp -Sim          & $-1.667 \; (.479)^{***}$ & $-2.577 \; (.316)^{***}$ & $-.344 \; (.126)^{**}$  \\
2Bar +Temp +Sim          & $-.930 \; (.380)^{*}$    & $-.261 \; (.466)$        & $.030 \; (.141)$        \\
4Bar -Temp +Sim          & $-1.193 \; (.372)^{**}$  & $-1.630 \; (.339)^{***}$ & $-.133 \; (.133)$       \\
4Bar +Temp -Sim          & $-1.930 \; (.488)^{***}$ & $-1.840 \; (.461)^{***}$ & $-.204 \; (.137)$       \\
4Bar -Temp -Sim          & $-1.457 \; (.470)^{**}$  & $-2.209 \; (.403)^{***}$ & $-.221 \; (.132)$       \\
4Bar +Temp +Sim          & $-1.720 \; (.505)^{***}$ & $-1.893 \; (.448)^{***}$ & $-.250 \; (.136)$       \\

\botrule
\end{tabular*}
\footnotetext{\scriptsize{$^{***}p<0.001$; $^{**}p<0.01$; $^{*}p<0.05$}\\}
\label{lmems_table}
\end{table}


\section{Discussion}\label{sec12}

In this study, we focused on human interaction with a piano-playing artificial partner. Instead of asking participants with prior piano experience to passively observe and evaluate the music produced by an AI, we asked them to play together with it in a turn-taking task. We then asked them to rate the quality of the interaction and we took measures of self-other integration and flow. This approach is in line with previous proposals for situated and embodied versions of the Turing test \citep{ortiz_why_2016,pfeiffer_non-verbal_2011,swisher_ascribing_2006} and for measures of emotional response and affiliation while coordinating with an artificial rhythmic stimulus \citep{kostrubiec_virtual_2015,launay_synchronising_2014,zhang_enhanced_2016}.

We found that the smallest model trained on 2-bar melodies and configured with high levels of imitation and either high or low levels of improvisation (2B,-T,+S and 2B,+T,+S) scored higher on measures of \textit{Realism} and \textit{Ease to Interact with}, approaching the scores of human partners. \textit{Self-Other Integration} in these same conditions was also similar to human partners. These are promising results as they indicate that the present approach with pre-trained generative models could enable an interactive turn-taking musical experience. Yet, all configurations of the artificial network were rated lower than the human partners on \textit{Enjoyment} and \textit{Creativity}. With respect to flow state, all conditions of the artificial partner and human partner scored similarly in the middle range of the scale. This suggests that the turn-taking task did not allow enough time and challenge for participants to settle in the sort of absorbed performance associated with flow.

During debriefing, participants shared that the artificial partner was naïve and limited but it was genuinely musical and interesting enough for them to use at home as a practice tool. In theory, the autoencoder is capable of exhibiting good next-step prediction. Participants observed that the artificial partner was trying to match and improvise from their short melodies but it was not able to continue and complete the ideas that they initiated; it was not sophisticated enough to re-create a rich musical call-and-response game. In brief, the results are promising but they also indicate that interactive performance is a difficult challenge for current frameworks in musical AI designed for passive imitation of piano performance.

{The objective of the study was to evaluate self-other merging and experience, not recognition of the partner as in the original Turing test. Features responsible for the differentiation of human- and machine-made sequences of sounds may not be the same as features that enable inter-personal interaction. For this reason, we did not take measures to blind participants with respect to the nature of their partner on each trial. As generative models advance in sophistication, however, it will be important to revisit experiments such as the classic Turing test. Using the apparatus of the current study, participants can be separated into individual rooms, unaware of the identity of their partners. Future research can address this and, additionally, expand the possibility for open-ended verbal feedback from the participants in the format of a qualitative interview to help understand in more details inter-personal interaction from their perspective.}

The implicit assumption in the present study was that before a machine could play music with a human, all relevant domain knowledge had to be squeezed into its software brain first, and then interaction would be solved. We assumed that the first stage had been achieved because a previous study by the designers of the MusicVAE found that the largest, most hierarchical model was capable of producing 16-bar sequences with musicality approaching human, as rated by listeners in a blind test \citep{roberts_hierarchical_2018}. In the present interactive test, however, the trend was the opposite, favoring short and more shallow models provided that they could respond immediately and adequately to the human partner. Future attempts could be more productive if they aim to design an AI directly for minimal interaction and then work on adding rich musical contexts, possibly from continued experience with human partners.

The present study points to a potentially productive use of artificial music partners as experimental apparatus in basic research on interpersonal coordination. The neural mechanisms involved in self-other segregation and integration during coordinated rhythmic action are only beginning to be explored \citep{heggli_kuramoto_2019,heggli_metastable_2021,liebermann-jordanidis_simultaneous_2021,novembre_neural_2016,palmer_are_2022,ragert_segregation_2014,zamm_behavioral_2021}. A common limitation in this context is the closure of free variables; there is no way to experimentally control the task as each participant stands for an interactive stimulus for the other participant. Inasmuch as simple rhythmic coordination is concerned, this problem can be avoided by using an interactive and parametrically controllable oscillator or another dynamic system, a so-called human dynamic clamp \citep{kostrubiec_virtual_2015}. Generalizing this research to music, however, requires richer and more realistic interactive partners. Such experimental artificial partners may be feasible with current tools for generative neural networks if they can be trained for interaction.

\subsection{Interactive AI}

\begin{quote}
\emph{"Our lives are not our own. From womb to tomb, we are bound to others, past and present, and by each crime and every kindness, we birth our future."} Sonmi-451
\end{quote}

The holy grail of research in AI has been to achieve strong AI, or artificial general intelligence (AGI), defined in terms of autonomy, generalization of learning, and understanding of the meaning and context in natural language, among others \citep{russell_artificial_2020}. The goal of developing fully autonomous vehicles generated waves of excitement in the last decade, yet the feasibility of this objective remains hotly debated and investment in the industry decreased sharply \citep{carey_focus_2023,glasner_self-driving_2022}. Neither are AI frameworks close to being able to use analogical reasoning to generalize their knowledge to novel domains \citep{mitchell_abstraction_2021}. Finally, the full set of requirements for social AI can be very challenging, including capacities such as communication of meaning and inference of the partner's intentional states \citep{fong_survey_2003}. 

In contrast, the conditions for inter-dependent human-machine problem solving are already present \citep{minsky_society_2013}. For example, it is a more realistic scenario to deploy vehicles with limited autonomy, expressed by the principle of human-in-the-loop control \citep{noauthor_dod_1998}. This is applied when neither a human operator nor an artificial expert system acting alone is capable of performing a given task as well as when the two are acting in collaboration \citep{minsky_society_2013}. Winograd and Flores \cite{winograd_understanding_1986} famously introduced the idea that the separation of subject (user) and object (machine) in human-computer interaction only appears in aberrant circumstances when fluid task performance is perturbed by a workspace malfunction. The case for the primacy of interaction over autonomous intelligence argues that interactive and social AI can be useful even if it is designed to be task-specific and inter-dependent with human partners \citep{froese_enactive_2009,minsky_society_1988,pfeifer_self-organization_2007,meyer_designing_1993,fong_survey_2003,bennett_emergent_2021,vicente_ecology_1990}. In what follows, we offer ideas on what it may take to design interactive AI.

At early stages of development, there needs to be greater emphasis on ability of the interactive AI to coordinate and synchronize with a human partner than on the ability to differentiate between high-level musical characteristics (i.e., the historical period and style of the musical piece). This is consistent with developmental trajectories in infants. From an early age, the development of inter-personal coordination skills and preferences for rhythmic style are honed by rhythmic interactions with their caregivers \citep{hannon_music_2007,trainor_rhythm_2015}. Needless to say, the evidence for a role of prediction-driven temporal expectations in cognition suggests that an artificial player with the ability to contextualize the partner’s stylistic and cultural preferences, without necessarily having comprehensive knowledge of every possible piece, may achieve better expectations of temporal variations \citep{hansen_predictive_2021,vuust_rhythmic_2014}.

A limitation of the generative model employed in the present study is that its architecture lacks recurrent loops between feed-forward passes. This means that it does not retain information between turns and, effectively, each turn is a separate trial from its perspective. Some chat bots may have the sophistication to retain all recent turns and use them as secondary input to further constrain their generated output. Yet, these architectures lack the ability to incorporate rhythmic musical timing. Arguably, in human interpersonal interaction, both the timing and the form of the response are crucial in enabling the sense of togetherness. {For the same reason, the turn-taking format is limiting because music making around the world typically involves people coordinating their playing at the same time. Designing generative models with an emphasis not only the content but also on its timing would require different network architectures, possibly incorporating recurrent neural networks optimized for dynamic synchronization. The ability to control synchronization parametrically will allow also to test whether the importance of imitation (similarity) observed here will generalize. Such a paradigm can be implemented with an exclusive focus on sound like here, or with an added visual modality by linking performance to an avatar with a body in virtual reality.}

For minimal interactive AI to be functional, it is sufficient that it enables coordination patterns in the shared space with the human partner; the pleasure of spontaneously falling in synchrony with someone else constitutes meaningful social experience regardless of the agency of the participants \citep{satne_shared_2020}. There are examples of modest artificial systems designed to induce spontaneous synchronization in the context of rhythmic behaviors. These usually have practical objectives such as the practice of social skills and rapport, improvement of gait and other motor function, or facilitation of musical performance \citep{dotov_entraining_2018,dotov_role_2019,nakata_perceptual_2015,raffard_using_2018}. These examples are rudimentary because they embody only one aspect of musical performance, namely entrainment of coupled oscillators by a pre-defined regular beat. Yet, such a constrained definition of interaction makes it possible to take advantage of progress made in related fields. Synchronization and control of dynamic systems has been investigated extensively in robotics, control theory, and applied dynamic systems theory. Recently, deep neural networks were used to learn the evolution of dynamic equations and extend the temporal window when predicting future states of chaotic systems \citep{lu_attractor_2018,pathak_model-free_2018}. Training artificial dynamic agents to synchronize and coordinate their musical performance with humans promises to reveal new horizons for social AI. Importantly, this implies that we need to focus our efforts not only on developing neural architectures but also on developing interactive paradigms for the behavioral training of AI.

Designing AI for interaction calls for special training principles. In the early days of connectionism, Geoffrey Hinton commented that multi-layer (deep) neural networks and unsupervised learning hold the potential to extract high-level patterns inherent in the stimulus space \citep{hinton_inferring_1980}, the approach assumed in the present work. Yet, training an artificial neural network to recognize a set of musical stimuli does not guarantee that it will be able to pick on invariants of coordination when playing with another musician, no matter how large the set is. The former is a classification task in an object-based ontology defined in the abstract space of musical excerpts. The latter is a real-time coordination task in a dynamic systems ontology defined over the joint space of multiple participants’ movements. This is more amenable to reinforcement learning with deep networks \citep{levine_learning_2018}, direct learning of affordances for coordination \citep{hasson_direct_2020}, or evolutionary algorithms \citep{kadihasanoglu_evolutionary_2017}.


\section{Conclusion}\label{sec13}

Variations of the famous Turing test can emphasize interactive and collaborative rather than generative capabilities of AI. This approach has better ecological validity given the inherent social nature of musical performance. For an artificial musical machine to emulate musicians so skillfully as to produce similar shared experiences, it would have to be designed and trained with interaction in mind. It remains to be seen if this is possible.

\backmatter

\bmhead{Acknowledgments}

We would like to thank Sally Stafford and Susan Marsh-Rollo for help with recruiting participants. This works was supported by a grant CIFAR awarded to LJT. DD received support from NIH P20GM109090 during preparation of this article.

\bibliography{ai_duo}

\end{document}